\def\year{2022}\relax
\documentclass[letterpaper]{article} 
\usepackage{aaai22}  
\usepackage{times}  
\usepackage{helvet}  
\usepackage{courier}  
\usepackage[hyphens]{url}  
\usepackage{graphicx} 
\usepackage{natbib}  
\usepackage{caption} 
\DeclareCaptionStyle{ruled}{labelfont=normalfont,labelsep=colon,strut=off} 
\usepackage{algorithm}
\usepackage{algorithmic}
\usepackage{amsmath}

\usepackage{xcolor}

\usepackage{newfloat}
\usepackage{listings}
\floatstyle{ruled}
\newfloat{listing}{tb}{lst}{}
\floatname{listing}{Listing}
\title{\textbf{Political Elites in the Attention Economy: Visibility Over Civility and Credibility?}}

\author{
    Ahana Biswas \textsuperscript{\rm 1}, 
    Yu-Ru Lin \textsuperscript{\rm 1}\footnote{Corresponding author.}, 
    Yuehong Cassandra Tai \textsuperscript{\rm 2},
    Bruce A. Desmarais \textsuperscript{\rm 2}
}
\affiliations{
    \textsuperscript{\rm 1} University of Pittsburgh\\
    \textsuperscript{\rm 2} Pennsylvania State University\\
    ahana.biswas@pitt.edu, 
    yurulin@pitt.edu, 
    yhcasstai@psu.edu,
    bdesmarais@psu.edu
}

\usepackage{booktabs}
\usepackage{threeparttable}

\usepackage{soul}

\def \yrv #1{{\color{black}{#1} }}
\def \yrvn #1{{\color{black}{#1} }}

\def \ahb #1{{\color{black}{#1} }}

\def \ahbf #1{{\color{black}{#1} }}

\begin{document}

\maketitle

\begin{abstract}
Elected officials have privileged roles in public communication. In contrast to national politicians, whose posting content is more likely to be closely scrutinized by a robust ecosystem of nationally focused media outlets, sub-national politicians are more likely to openly disseminate harmful content with limited media scrutiny. In this paper, we analyze the factors that explain the online visibility of over 6.5K unique state legislators in the US and how their visibility might be impacted by posting low-credibility or uncivil content. We conducted a study of posting on Twitter and Facebook (FB) during 2020-21 to analyze how legislators engage with users on these platforms. The results indicate that distributing content with low-credibility information attracts greater attention from users on FB and Twitter for Republicans. Conversely, posting content that is considered uncivil on Twitter receives less attention. A noticeable scarcity of posts containing uncivil content was observed on FB, which may be attributed to the different communication patterns of legislators on these platforms. In most cases, the effect is more pronounced among the most ideologically extreme legislators. Our research explores the influence exerted by state legislators on online political conversations, with Twitter and FB serving as case studies. Furthermore, it sheds light on the differences in the conduct of political actors on these platforms. This study contributes to a better understanding of the role that political figures play in shaping online political discourse.

\end{abstract}
\section{Introduction}
Social media channels are effective for communication due to the ease of information dissemination irrespective of the quality of the information. Politicians also leverage social media due to the wide reach of these platforms. While public officials can promote constructive dialogue, they can also spread harmful content online. Government officials are often subjected to less stringent content moderation rules \cite{pelletier2021fexit}, and have higher followings than average users~\cite{grant2010digital} which makes it easier for them to endorse and propagate harmful content online. 

Political communication online is often geared toward the target audience and platform characteristics~\cite{kreiss2016seizing,enli2013personalized,stier2020election, kelm2020politicians}. 
For instance, \citet{stier2020election} found that social media messages of both candidates and their audiences focused on distinct topics compared to the general audience during 2013 German federal elections. More broadly, politicians often tailor their content to achieve maximum political gains, and prior studies have shown that social media plays a significant role in shaping political agendas, influencing public opinion, and potentially affecting electoral outcomes ~\cite{kreiss2016seizing,bovet2019influence,boulianne2023engagement}.


To understand how politicians use social media to sway public opinion, it is critical to examine what makes them visible online. \ahb{In this study, we use the term visibility to refer to the level of public engagement or interactions individuals receive on social media, which is often utilized to gauge their outreach and influence on these platforms}\ahbf{~\cite{eberl2020s}}. Party membership, demographic information (such as state, race, and gender), and platform-level characteristics (such as follower count, posting activity, and content style) are potential contributors. We are particularly interested in determining if politicians' visibility is increased by the dissemination of {\it harmful} content, which includes toxic and uncivil language, as well as untrustworthy or non-credible information. Both uncivil and low-credibility content have been associated with a decline in the quality of democratic discourse~\cite{goovaerts2020uncivil,bennett2018disinformation}. We ask, \textit{does posting uncivil and low-credibility content increase the visibility of politicians?} The answer to this question is critical as prior research has linked harmful content online to violent offline incidents, increased affective polarization, distrust in institutions, and so on~\cite{johnson2018self,serrano2021digital,coe2014online}. Harmful content originating from or endorsed by politicians may further exacerbate these negative outcomes due to their larger audience base and higher trustability owing to partisan preferences.

This work focuses on how US state legislators cultivate and exert their influence online. In contrast to national politicians, who are closely scrutinized by media outlets \cite{kyriakidou2021journalistic}, sub-national politicians are more likely to disseminate harmful content with limited scrutiny~\cite{mihailidis2021cost}. State legislators are responsible for laws across all policy areas within state jurisdiction, making their role crucial in the U.S. political system. Given the limited media coverage of state legislators~\cite{squire2019state}, these social media platforms serve as important mediums for communicating their ideological and political positions to their voters.  

We study the dynamics of legislators’ visibility, examining the different factors that influenced the attention they received on Twitter and Facebook (FB) during the two-year period spanning 2020-2021. We focus on these two years owing to the surge of harmful content online due to significant events such as the US Presidential elections, COVID-19, Capitol Riots, and BLM protest movements~\cite{ferrara2020characterizing,cuan2020misinformation,toraman2022blacklivesmatter}. \ahb{Studying the visibility dynamics over a time period has certain challenges. Apart from individual attributes (e.g., posting frequency, party, demographics) or volume of harmful content posted, the politician's visibility may also vary by time (e.g., during elections) or due to the particular topics they post. We tackle these challenges in our study, our main contributions are as follows}:

\begin{itemize}
    \item \ahb{\textbf{Factors Associated with Visibility.}} We present a large-scale, longitudinal study on political elites' online visibility in the US \ahb{by comparing} differences in their platform visibility based on party, socio-demographic factors, and posting activity \ahb{(RQ1, RQ2; See Section 3)}. Republicans and men tend to have a higher level of visibility on FB, while Democrats tend to have higher visibility on Twitter. Posting uncivil content on Twitter and similarly low-credibility content on FB is also correlated with their platform visibility. Legislators' visibility on posting low-credibility content, however, varies by party on Twitter. Our thorough analysis of legislators who post on both platforms reveals notable platform differences associated with their social media activities.




    \item \ahb{\textbf{Methodological Contribution.}} We conduct a causal inference study to examine how legislators' social media posts affect their visibility, particularly when the content is uncivil or less credible \ahb{(RQ3; See Section 3)}. To ensure that our findings are not influenced by potential confounding factors, such as temporal and topical correlations that are common in dynamic text data and can bias the results, we leverage deep learning of potential outcome and matching techniques. Our analytical method helps disambiguate the effect of posting activities.

    \item \ahb{\textbf{Impact of Harmful Content.}} The results, based on observational data using causal inference \ahb{(RQ3; See Section 3)}, have revealed significant and novel patterns. Our study found that posting uncivil content on Twitter led to a decrease in visibility. It was observed that Republicans posting low-credibility content on both platforms have an increased visibility, while Democrats posting the same have lower visibility. The effect is more pronounced for ideologically extreme legislators in most cases. Overall, our findings contribute to the understanding of politicians' online visibility, shedding light on cross-platform differences and partisan asymmetries.
    
\end{itemize}

\section{Related Work}

\paragraph{Political Elites' Online Behaviors.}
Social media are used by politicians for both broadcasting as well as having dialogue with voters. The effect of Twitter and Facebook use on election campaigns has been studied extensively~\cite{kreiss2016seizing,jungherr2016twitter,boulianne2023engagement, sahly2019social}. \citet{kreiss2016seizing} looked at how Twitter was used by political party staffers to shape the perspectives of journalists and influence dedicated voters. Voters engaging in political discussion online have demonstrated increased interest and engagement in political affairs~\cite{bode2016politics}.
Social media, thus, serves as a powerful tool for politicians to influence public opinion and/or convey their stance regarding several important issues. \ahb{Despite a large body of work on political communication on social media, there is no clear understanding of which factors influence the online visibility of legislators, especially, how politicians posting \textit{harmful} content is viewed by the audience. Our research aims to close this gap by examining the impact of \textit{uncivil} and \textit{low-credibility} content on legislators' visibility by performing a cross-platform study---after accounting for several confounding factors related to their personal attributes, temporal and topical variations.} 



\paragraph{Misinformation and Virality.}

There exists a large body of literature characterizing the diffusion of low-credibility content online~\cite{vosoughi2018spread,friggeri2014rumor,zollo2015emotional}. \citet{vosoughi2018spread} found that falsehoods spread significantly faster, and reached a broader audience as compared to true news on Twitter. \citet{friggeri2014rumor} found that rumor cascades on FB tend to penetrate deeper into the social network compared to general reshare cascades. Prior works suggest that the sentiment towards misinformation is primarily negative which could be responsible for the variations seen in the diffusion~\cite{vosoughi2018spread,zollo2015emotional}.

A significant body of research has examined the impact of misinformation on the 2016 and 2020 US presidential elections~\cite{bovet2019influence,pennycook2021examining}. 
Prior research has also shown that online misinformation tends to be directed more frequently toward conservative users~\cite{rao2022partisan, yang2023visual} making them more likely to engage in misinformation. Misinformation may have a higher reach on social media platforms which could be leveraged by politicians to gain visibility, and the extent may vary across ideologies. \ahb{However, the question of how misinformation originating from public figures is reacted by audiences is less explored. In this work, we aim to illuminate the impact of posting low-credibility content on legislators' visibility.}

\paragraph{The Attention Economy and Toxic or Controversial Content.}

There is no clear consensus on how uncivil content spreads on online platforms~\cite{shmargad2022social,gervais2015incivility}. Prior works have studied the nature of incivility in online political communication suggesting that engaging in uncivil discourse may have certain benefits for politicians such as political opinion polarization~\cite{bodrunova2021self} or empowerment by voicing criticism against authorities~\cite{bodrunova2021constructive}. 
Uncivil content was found to be associated with emotionally loaded language which generated strong responses from the audiences~\cite{mutz2007effects}. Irrespective of the kind of response, this may lead to higher visibility. \citet{coe2014online} found that uncivil comments on news websites received more negative reactions. 
\ahb{Thus, even though engaging in uncivil discourse may have certain political benefits, it is unclear how that is perceived by audiences---a gap that we address in this study.}


\paragraph{Confounding with Textual Data.} Causal inference with text data is particularly challenging since the assumptions of causal inference (positivity, conditional ignorability, consistency) may not hold when confounding, treatment or outcomes are encoded in text~\cite{feder2022causal}. For instance, posts on certain topics may be more likely to contain misinformation and also receive higher engagement from the audience. Prior works on extracting confounding from text have utilized unsupervised dimensionality reduction methods~\cite{roberts2020adjusting,sridhar2019estimating}. Recent works leverage neural networks to automatically extract features especially when the confounders in text are not explicitly known~\cite{koch_sainburg_geraldo_jiang_sun_foster_2021}. To address this problem, some works have added transformer layers for text processing to TARNet or Dragonnet~\cite{veitch2020adapting,pryzant2020causal}. \ahb{We have extended the state-of-the-art techniques to address the confounding factors that are commonly seen in dynamic social media content, such as textual and temporal correlations due to similar topics, events, and personal attributes. To generate content representations, we leverage contextual RoBERTa embeddings~\cite{liu2019roberta} with other post attributes. We then utilize a fine-tuned Dragonnet model to produce content embeddings that isolate the confounding factors.}


 
\section{Study Design}
It is crucial to understand how politicians develop influence through online media and factors associated with the influence.  This work focuses on state legislators specifically since they may be more likely than national-level politicians to disseminate harmful content owing to limited scrutiny. We ask the following research questions (RQs):

\begin{itemize}
\item[{\bf RQ1}] How does the legislators' visibility, as measured by the attention they receive, vary based on party affiliation, and individual attributes such as gender, ethnicity, state representation, and social media activity?
\item[{\bf RQ2}] What attributes of legislators and their posts are associated with their visibility?
\item[{\bf RQ3}] How does low-credibility or incivility impact the visibility of legislators' posts?
\end{itemize}

In RQ1 we analyze whether the attention received by legislators varies based on party affiliation, basic demographics, and posting activity. RQ2 examines what characteristics of legislators are most strongly correlated with their online visibility change. RQ1 and RQ2 characterize how visibility varies by legislators' attributes and provide an understanding of factors that may potentially impact their visibility dynamics at the account level. To examine the impact of posting harmful content, RQ3 investigates how low-credibility and incivility {\it impacts} the attention received by individual posts. More specifically, we study whether low-credibility or incivility increases or decreases a post's visibility where visibility is measured in terms of expected interaction rate. 
Since the user attention on social media platforms may vary by legislators' attributes (RQ1), and there may be other factors associated with the visibility (RQ2), in RQ3 we estimate the impact of incivility or low-credibility by controlling for these variables as well as temporal and topical variations.

\subsection{Datasets} \label{data}
We collect Twitter and FB posts from all US state representatives and senators who held office during \ahb{2020-2021} \ahb{(i.e., each legislator has been in office at least for a certain time between 2020 and 2021)}. We focus on Twitter and FB due to the vast amounts\footnote{https://www.pewresearch.org/internet/2020/07/16/congress-soars-to-new-heights-on-social-media/} of content produced by legislators on these platforms. Of the 8,028 legislators, 5,712 (64\%) legislators, comprising 2,943 (61\%) Democrats, 2,740 (48.2\%) Republicans, and 29 Independents had at least one Twitter account~\cite{kim2022attention}. For FB, 5,147\footnote{FB account information was crawled from Ballotpedia} (64.1\%) legislators, comprising 2,215 (48.2\%) Democrats, 2,515 (56.0\%) Republicans, and 418 other party members, had either an official or a campaign account. We collect all their Twitter and FB posts during 2020-21. \ahb{It is important to note that many of these accounts were either dormant or inaccessible during our data collection period\footnote{See Appendix for further details on data collection}.}

\noindent \textbf{Twitter.} Our Twitter dataset\footnote{Collected using Twitter API v2 before March 2023} comprises around 4M tweets posted by 3,551 (44.2\%) US state legislators during 2020-2021. The coverage of state legislators ranges from 29.4 to 96.5\%, with a mean of 71.6\%. This suggests that our dataset comprises a representative legislator population across most US states. We only analyze tweets by Democrats and Republicans due to the insufficient number of posts from Independent legislators (N=29). We calculate the interaction a tweet receives as the sum of likes, retweets, replies, and quotes. Table~\ref{tab:data_stats} shows the number of legislators, posts, and median\footnote{The interactions received on posts have a skewed distribution, so we report the median instead of mean.} interactions on post. Democrats are more active and receive higher engagement on posts as compared to Republicans. The comparatively higher volume of posts by Democrats indicates that Twitter is a more preferred communication platform for them compared to Republicans. We also construct the intra-legislator follower network, where a directed edge from legislator $A \rightarrow B$ indicates $A$ follows $B$.\\

\noindent \textbf{FB.} We collect all FB posts\footnote{Collected using CrowdTangle API} by US state legislators during 2020-21. This yields over 493K posts from 5,147 (64.1\%) legislators. The coverage of state legislators ranges from 23.5 to 80.6\%, with a mean of 57.7\%. We similarly focus only on posts from Republicans and Democrats due to a small number of posts from other party members. We use the interaction metric returned by the CrowdTangle API which is the sum of all reactions (`Likes', `Love', `Wow', `Haha', `Sad', `Angry', `Care'), comments, and shares for a post. Table~\ref{tab:data_stats} shows the number of legislators, posts, and median interactions on post. The posting frequency is similar for Democrats and Republicans on FB unlike on Twitter. Interestingly, Republicans receive almost double the interaction as compared to Democrats, suggesting that Republicans may have a larger audience base and hence a larger reach on FB. We do not use the follower count for FB data since some of the legislator accounts are official accounts while others are Pages, due to which the follower counts across these different account types vary widely. 
\yrv{We are unable to analyze FB's network data due to the lack of access.}



\begin{table}
\centering
\setlength{\tabcolsep}{3pt}
\caption{Descriptive statistics for Twitter and FB datasets showing the number of legislators, posts, and median interactions received per post by party.}\label{tab:data_stats}
\small
\begin{tabular}{lccccccc}
\toprule
 & \multicolumn{3}{c}{\textbf{Twitter}} & \multicolumn{3}{c}{\textbf{FB}} \\
\cmidrule(rl){2-4} \cmidrule(l){5-7}
\textbf{party} & \textbf{\#users} & \textbf{\#tweets} & \textbf{Int} & \textbf{\#users} & \textbf{\#posts} & \textbf{Int}\\
\cmidrule(r){1-1} \cmidrule(rl){2-4} \cmidrule(l){5-7}
Dem & 1677 & 2.25M & 8.0 & 2211 & 224K & 58.0\\
Rep & 1412 & 889K & 6.0 & 2501 & 218K & 101.0\\

\bottomrule
\end{tabular}\vspace{-1em}
\end{table}

\subsection{Individuals' attributes}\label{sec:att}
We characterize legislators based on their platform presence and individual-level characteristics. For Twitter-based attributes, we include their post count, follower count, and in-degree centrality in the intra-legislator follower network. The post count serves as a proxy for measuring how actively the legislators use the platform and follower count indicates the size of their audience base. Both post and follower counts are likely to have an impact on online visibility~\cite{hasan2022impact}. Legislator's position in their peer network may also impact their visibility (larger follower base, greater potential for content virality, or seniority). To gauge the influence of the legislators among their peers concerning the immediate connections, we leverage the in-degree centrality measure. For FB-based attributes, we include the post count. The individual-level attributes include party affiliation (Republican vs. Democratic), state, gender (Men vs. Women), ethnicity (White vs. Non-White), and ideology scores. We leverage the ideology scores constructed by \citet{shor2011ideological}. Around 99.7\% (N=3074) and \ahb{70.2\%\footnote{\ahb{See Data Collection section in Appendix}} (N=3306)} of legislators are mapped\footnote{Ethnicity and gender are mapped using Ballotpedia. Binary genders are used due to insufficient data about non-binary genders.} to their attributes on Twitter and FB, \ahb{and only those legislators with attributes are analyzed in our study}. The final dataset comprises around 62.2\% and 53.8\% White, and 68.2\% and 67.7\% men on Twitter and FB respectively, indicating that men and White legislators outnumber women and Non-Whites respectively. There are 2,131 (26.5\%) overlapping users (OL) \ahb{(i.e., legislators having accounts on both Twitter and FB)} across Twitter and FB. Figure~\ref{fig:att} \ahb{in Appendix} shows the breakdown of ethnicity and gender by party and for OL. Republicans have fewer women users compared to Democrats for both Twitter and FB and both platforms have more Republican men. The representation of Non-White Republicans is higher on FB than on Twitter.


\section{Measuring Posts' Civility, Credibility, and Legislators' Visibility}

\subsection{Assessing posts' civility}

\ahb{We assess the civility of a post based on the toxicity of their language}\ahbf{ which is a common practice in literature~\cite{frimer2023incivility,kim2021distorting}.} \ahb{Incivility in the context of political speech is commonly associated with rudeness according to the study by~\citet{stryker2016political}. We follow the definition of \textit{toxic} language provided by Google Perspective\footnote{https://perspectiveapi.com/how-it-works/}: \textit{``rude, disrespectful, or unreasonable comment that is likely to make someone leave discussion''}. Based on this definition, it would be reasonable to assume that the toxicity classifier is able to identify the markers of political incivility.}  We determine the level of toxicity using the ``original'' model\footnote{The ``original'' model had the best performance when evaluated against our manual annotation labels compared to the other Detoxify models, namely, ``debiased'' and ``multilingual''.} from Detoxify\footnote{\ahb{We choose Detoxify over Google Perspective API since Detoxify has better or comparable performance~\cite{arhin2021ground} and is faster}} \cite{Detoxify}. We choose the cutoff for toxicity score as 0.82 based on manual annotation (see 
Appendix), i.e., posts having a score above 0.82  are considered uncivil\footnote{See Appendix for examples of uncivil posts}.
This yields 24,242 (0.8\%) and  277 \ahb{(See Appendix)} uncivil posts on Twitter and FB. 

Table~\ref{tab:tox_stats} shows the number of legislators posting uncivil content, posts, and median interactions received, by party. This may be attributed to politicians' different communication styles across these platforms as observed in prior research, suggesting FB is used more for broadcasting purposes compared to Twitter which is leveraged more for having dialogue~\cite{enli2013personalized}. For our analysis on uncivil content in the following sections, we only focus on Twitter owing to the small number of uncivil posts on FB. Around 47.1\% of Democrats and 32.6\% Republican legislators post uncivil content on Twitter. Democrats post almost double the number of uncivil content compared to Republicans (on average) on Twitter. This could be due to the higher interaction received on such posts by Democratic legislators. The interaction on uncivil content is higher compared to the baseline interaction (in Table~\ref{tab:data_stats}) for both parties, with a larger difference for Democrats. 

\begin{figure}[ht]
\centering
\includegraphics[width=\linewidth]{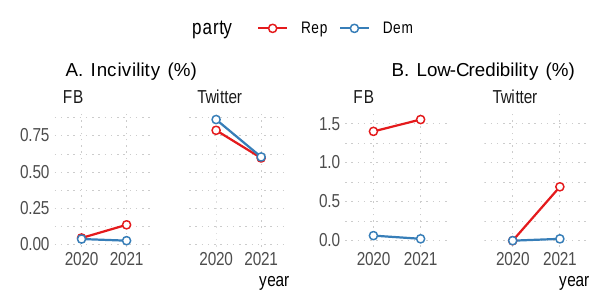}
\caption{{\bf Percentage of uncivil and low-credibility posts across years, by party, by platform.} Republicans have a higher rate of posting low-credibility content on both platforms and across years. The yearly posting rates of uncivil content are comparable across parties.} \label{fig:year_party}
\end{figure}

\begin{figure}[ht]
\centering
\includegraphics[width=\linewidth]{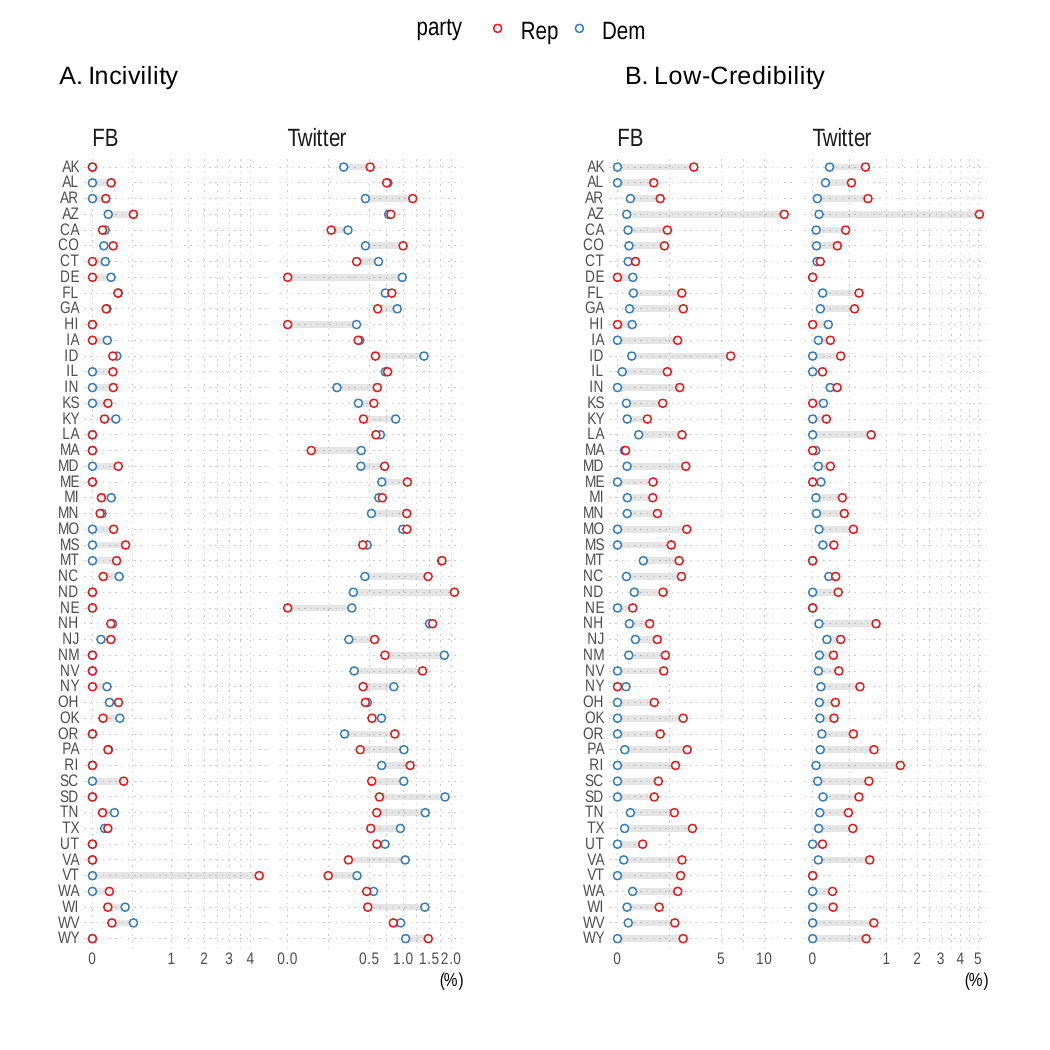}
\caption{{\bf Percentage of uncivil and low-credibility posts from states, by party, by platform.} Republicans have a higher rate of posting low-credibility content across all states, on both platforms. Moreover, Republican legislators from most states have a higher rate of posting uncivil content on FB. State and party-wise differences exist for posting rates of uncivil content on Twitter. Note that the x-axis scale has been adjusted to better visualize party differences.} \label{fig:state_party}
\end{figure}

Figure~\ref{fig:year_party}A shows the rate of uncivil posts across years by party and platform. More uncivil content was posted during 2020 on Twitter, with Democrats having a higher rate of posting uncivil content. Figure~\ref{fig:state_party}A shows the rate of uncivil posts across states by party and by platform. Interestingly, we find that Republicans posted more uncivil content on FB compared to Democrats across almost all the states, but state-wise differences exist for Twitter.

\begin{table}
\centering
\setlength{\tabcolsep}{3pt}
\caption{Number of uncivil posts, legislators posting, and median interactions received per post on uncivil content on Twitter and FB, by party.}\label{tab:tox_stats}
\small
\begin{tabular}{lccccccc}
\toprule
 & \multicolumn{3}{c}{\textbf{Twitter}} & \multicolumn{3}{c}{\textbf{FB}} \\
\cmidrule(rl){2-4} \cmidrule(l){5-7}
\textbf{party} & \textbf{\#users} & \textbf{\#tweets (\%)} & \textbf{Int} & \textbf{\#users} & \textbf{\#posts (\%)} & \textbf{Int}\\
\cmidrule(r){1-1} \cmidrule(rl){2-4} \cmidrule(l){5-7}
Dem & 782 & 18111 \ahb{(0.9\%)} & 14.0 & 55 & 89 \ahb{(0.1\%)} & 131\\
Rep & 461 & 6131 \ahb{(0.7\%)} & 7.0 & 54 & 188 \ahb{(0.1\%)} & 120\\

\bottomrule
\end{tabular}\vspace{-1em}
\end{table}

\subsection{Assessing posts' credibility}

We identify low-credibility content based on the credibility of URL in the post, which is a common practice in literature~\cite{lasser2022social}. \ahb{It is important to note that this method of identifying low-credibility posts does not discriminate between posts endorsing or debunking such information. In this study, we are interested in looking at the visibility of both, i.e., posts promoting or debunking low-credibility information since both types of posts are exposing the audience to harmful information. Prior research has shown that exposure to misinformation has an impact on believing and subsequent sharing of such information~\cite{halpern2019belief}. Thus, irrespective of the author's intent, the audience may be susceptible to believing and/or sharing such low-credibility content. Moreover, the URL sharing patterns are similar for Democrats and Republicans on both Twitter and FB\footnote{Around 19.2\% posts by Republicans and 17.4\% by Democrats contain URLs on Twitter, and 14.6\% and 18.3\% posts by Republicans and Democrats on FB contain URLs.}, so it is unlikely that any biases are introduced in the study due to our method of using URLs to identify low-credibility posts.} We use the low-fact URL domain references provided by \citet{tai2023official} to identify low-credibility posts. \citet{tai2023official} refined  Media
Bias/Fact Check’s\footnote{https://mediabiasfactcheck.com/} (MBFC) original ratings to include URLs that contain misleading information and not just politically biased ones. This restrictive framing may undermine the true scale of misinformation on these platforms but it offers a higher precision (vs. recall) in identifying low-credibility content. This yields 6,848 (0.2\%) and 4,141 (1.0\%) low-credibility posts on Twitter and FB respectively suggesting that legislators post more low-credibility content on FB. 

Table~\ref{tab:mis_stats} shows the number of legislators posting low-credibility content, posts, and median interactions received, by party. Around 5.2\% of Democrats and 36.7\% Republican legislators post low-credibility on FB. On Twitter, only a handful of accounts are responsible for spreading low-credibility content across both parties (1.1\% for Democrats and 2.8\% for Republicans). Republicans post more low-credibility content compared to Democrats on both platforms. Similar to uncivil content, posts containing low-fact URLs receive higher interaction except for Democrats on Twitter.  The median interaction for low-credibility content is almost three times on Twitter and double on FB for Republicans compared to their baseline interaction\footnote{\ahb{For Twitter, the median interaction on posts with and without URLs are 9.0 and 7.0 respectively whereas for FB, the median interactions are 62.0 and 81.0, i.e., there is no clear pattern as to whether having a URL increases or decreases the interaction of posts, suggesting that the results in Table 3 could potentially be due to the low-credibility of URLs. }}.

Figure~\ref{fig:year_party}B shows the rate of low-credibility posts across years by party and platform. The prevalence was higher during 2021 on both platforms and it was mainly driven by Republicans. Figure~\ref{fig:state_party}B shows the rate of low-credibility posts across states by party and by platform. Low-credibility content is driven by Republicans across all the states,  with the highest rate from Arizona, on both Twitter and FB. \ahb{The low-credibility information from Arizona is mostly related to the 2020 US Presidential elections. In particular, we find that some Republican legislators frequently shared posts from unreliable information outlets, especially during Arizona's audit of the 2020 election, which contributed to a significant amount of low-credibility posts from Arizona.}

\begin{table}
\centering
\setlength{\tabcolsep}{3pt}
\caption{Number of posts containing low-fact URLs,  legislators posting, and median interactions received per post on low-credibility content on Twitter and FB, by party.}\label{tab:mis_stats}
\small
\begin{tabular}{lccccccc}
\toprule
 & \multicolumn{3}{c}{\textbf{Twitter}} & \multicolumn{3}{c}{\textbf{FB}} \\
\cmidrule(rl){2-4} \cmidrule(l){5-7}
\textbf{party} & \textbf{\#users} & \textbf{\#tweets (\%)} & \textbf{Int} & \textbf{\#users} & \textbf{\#posts (\%)} & \textbf{Int}\\
\cmidrule(r){1-1} \cmidrule(rl){2-4} \cmidrule(l){5-7}
Dem & 19 & 188 \ahb{(0.0\%)} & 4.0 & 83 & 114 \ahb{(0.1\%)} & 78.5\\
Rep & 40 & 6660 \ahb{(0.8\%)} & 20.0 & 567 & 4027 \ahb{(2.6\%)} & 239.0\\

\bottomrule
\end{tabular}\vspace{-1em}
\end{table}

The differences in interactions received by low-credibility and uncivil posts may be attributed to the differences in post topics, timing, or attributes of authors---we address this in the following sections.

\subsection{Measuring legislator's visibility} \label{vis_meas}

\yrv{Social media is being increasingly used as a tool by politicians to enhance their visibility and outreach to the public}\ahbf{~\cite{bahramirad2022virtual}.} \yrv{The measure of a legislator's visibility is typically based on the interactions received on their posts. Our approach to measuring visibility is inspired by the metrics used on Twitter and Facebook to calculate a post's expected engagement or virality potential~\cite{twitter,fb}. Our objective is to measure the overall outreach of a post or legislator on the platform. Therefore, we consider all the interactions received on a post or by legislators instead of focusing on a single type of engagement such as ``Likes.'' It's important to note that the visibility metrics are platform-specific and may not be comparable across platforms even if they share the same names. For instance, audiences may engage with ``Like'' feature on Twitter differently than ``Like'' on Facebook due to different interface designs (see details in Appendix). However, our visibility metrics are designed to capture the overall level of engagement a post or legislator receives on a specific platform.}

Moreover, we do not distinguish between positive and negative reactions received from the audience (e.g., `love' vs. `angry' on FB). We are concerned with the visibility of the legislators and both positive and negative reactions contribute to their overall outreach on the platform. As shown in Tables~\ref{tab:tox_stats} and~\ref{tab:mis_stats}, the interactions received on low-credibility and uncivil content are noticeably different from the overall interactions received by legislators, suggesting that audiences' reactions differ based on content (or other related factors). Thus, using interactions as a proxy to measure legislators' online visibility allows us to capture these differences and get insights into factors contributing to their visibility. 

The visibility can be measured at both account and post level. For post level, the visibility of $i^{th}$ post by legislator $u$, $v_{ui}$ is simply the interaction received on that post. Furthermore, we examine how visibility {\it changes} in relation to other variables. Instead of measuring aggregated interactions over the audience of their posts, we measure the legislator's visibility ($V$) by interaction rate, i.e., the number of interactions per post or audience size. We consider three different ways to adjust the quantity of the interaction rate, since platform reach tends to be correlated with the number of followers~\cite{hasan2022impact}, resulting in the following \textit{dependent variables} (DVs): (1) interaction normalized by post count ($V^{I_P}$), (2) interaction normalized by follower count ($V^{I_F}$), and (3) interaction normalized by follower and post count ($V^{I_{P, F}}$). 
Therefore, the visibility of legislator $u$, $V_u$, is given by aggregating the visibility of all $u$'s posts normalized by post and/or follower count.

\section{Methods}\label{methods}
In this section, we describe the methods used to answer our RQs. In addition to the legislator behavior on Twitter and FB, we analyze the overlapping legislators (OL) to understand whether the differences observed across these platforms are due to different legislator populations or different audience/platform characteristics on Twitter and FB. 

\subsection{Analyzing Legislators' Visibility}
In RQ1, we analyze whether legislators' visibility varies based on demographics, party, and posting activity. For attention disparity based on posting activity, we compare visibility of legislators having above the median number of posts with those posting less than or equal to median\footnote{We chose median as the threshold due to the skewed posting activity distribution.}. Using Mann-Whitney U test, we find differences in the platform visibility of legislators based on party, gender, ethnicity, and posting frequency. We incorporate additional DV variants, namely, 25$th$, 50$th$, and 75$th$ percentile of the legislator's post interactions together with the mean interaction, i.e., $V^{I_P}$ to ensure robustness of our results. For states-wise differences, we visualize the mean visibility across states.

For RQ2, we study the factors significantly correlated with the platform visibility of legislators. In addition to the individuals' attributes (Section~\ref{sec:att}), we measure the association between low-credibility and uncivil post volumes and their visibility. Since low-credibility content is targeted more towards conservative users~\cite{rao2022partisan,yang2023visual}, we also consider the interaction between party and low-credibility content posted. Additionally, the legislator's visibility may be influenced by the visibility of their peers if their peers (re)post similar content often. To measure this network visibility, we use the median of the visibilities of legislator's in-degree neighbors in the intra-legislator follower network. Unfortunately, we are unable to account for the network effects on FB.
We use the following model,

\begin{equation} \label{reg1}
\begin{split}
V_u &= \beta_0 + \beta_PParty_u + \beta_GGender_u + \beta_EEthnicity_u + \\ & \beta_NPosts_u + \beta_FFollowers_u + \beta_CCentrality_u + \\ 
& \beta_SState_u + \beta_TUncivil_u + \ahb{\beta_MLow Credible_u} + \\ & \beta_{Net}NV_u + \beta_{e}Party_u * \ahb{Low Credible_u}
\end{split}
\end{equation}

where $V_u$ is the visibility of legislator $u$, $Uncivil_u$ and \ahb{$Low Credible_u$} are the count of uncivil and low-credibility posts, $Party_u, Gender_u, Ethnicity_u$ and $State_u$ are dummy variables, $Posts_u$ is the post count, $Followers_u$ is the follower count\footnote{Only included for DV not normalized by follower count} (Twitter specific), $Centrality_u$ is the indegree centrality in intra-legislator follower network (Twitter specific), and $NV_u$ is the network visibility. To address the correlated errors across and within states, we incorporate a random effect on the state variable. The effect of each factor is estimated using a linear mixed effects regression model\footnote{To satisfy assumptions of linear regression, all variables are suitably transformed to be close to normal distributions (See Appendix). Ideology score is dropped since it is correlated with party.} \ahb{with standardized continuous variables.} 
\yrv{To ensure robust results for the analysis of RQ1 and RQ2, we conduct separate analyses for the years 2020 and 2021, to determine if the visibility trends observed were consistent across both years\footnote{Our findings revealed that the trends were similar for both years. Therefore, we have reported the overall results for the entire study period.}. Moreover, we analyze the topical\footnote{Identified using keywords.} distributions (e.g., COVID-19, BLM, elections) for our dataset and find that our data is not skewed towards any particular topic, suggesting limited bias due to specific topic(s).}


\subsection{Analyzing the Impact of Low-credibility or Uncivil Content}

\subsubsection{Measuring outcome.}
Posting low-credibility and uncivil content may have an impact on how politicians are perceived online. In particular, we analyze whether the presence of incivility or low-fact URLs increases/decreases their posts' visibility. To characterize the change in visibility, we analyze the engagement on a post considering the post author's expected visibility. Our metric is inspired by the overperforming metric used at CrowdTangle~\cite{fb}. The overperforming score for post $i$ is calculated as follows,

\begin{equation}
    O_{ui_p} = \frac{v_{ui_p}}{b_{u_p} + thres_p}
\end{equation}

where $b_u$ is the median interaction for legislator $u$'s posts on the platform in previous $w$-days and a threshold ($thres$) for the minimum number of interactions on a post to be considered as overperforming, with $p$ denoting platform. The term $b_{u_p}$ is used to adjust the outcome with respect to the level of expected interactions with the post authors on platform $p$.
We choose the $thres=10$ for Twitter (i.e., a post must have at least 10 interactions to be considered overperforming on Twitter) and $100$ for FB. We estimate the ideal window $w$, based on legislators' daily posting rates on these platforms (see Appendix for \ahb{$thres$}, $w$ estimation). To get a reliable estimate of the overperforming metric, we choose $w=14$ day rolling window. This allows us to calculate the overperforming score over a sufficient number of posts per legislator, while also accounting for the temporal variation. 

\begin{figure}[ht]
\centering
\includegraphics[width=\linewidth]{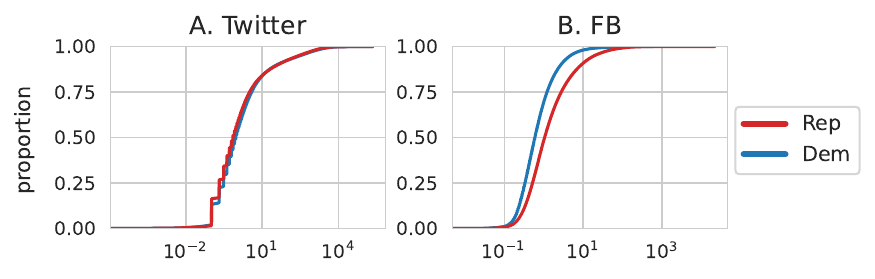}
\caption{{\bf ECDF plots of overperforming score distributions by party, by platform.} Distributions are similar for Twitter but posts by Republicans overperform more on FB.} \label{fig:ov}
\end{figure}

Figure~\ref{fig:ov} shows the post overperforming scores on Twitter and FB\footnote{Our scores for FB posts are highly correlated (Spearman rho=0.997, $p<0.001$) with the overperforming score returned by the CrowdTangle API, suggesting that our method successfully identifies posts that are overperforming.}. The distribution of overperforming scores are similar for Republicans and Democrats on Twitter. On FB, posts by Republican legislators overperform more compared to posts by Democrats. This suggests that posts by Republican legislators have a higher tendency to be viral on FB compared to Democrats. For estimating the causal impact, we are interested in analyzing whether a post overperforms or not due to its incivility or low-credibility, hence we binarize the overperforming score ({\it outcome})  at cutoff $1$, i.e., a post is overperforming if it has a score $>1$.


\subsubsection{Estimating causal impact.}
It is necessary to control for the \textit{confounders} that affect both the \textit{treatment} and \textit{outcome} variables to differentiate between {\it spurious correlation} and {\it causation}. One of the possible confounders is the topic---posts on certain topics (e.g., COVID-19) may be more likely to contain misinformation, and these topics may get higher visibility. Another possible confounder could be the tone---posts having certain tones (e.g., assertive) may be more likely to contain uncivil language, and also more likely to generate stronger responses from the audience. Apart from textual content, other confounding variables can include legislators' personal traits and the timing of their posts (e.g., during elections there might be a rise in harmful content as well as an increase in legislators' visibility). 

We control for the individual's attributes mentioned in Section~\ref{sec:att} (excluding ideological scores), post content, and time, i.e., count of days since the content was posted, starting from 2020-01-01. To control for the content, we leverage embeddings generated by the pre-trained RoBERTa model. For low-credibility content, we also include the URL headlines along with post content because we assume that both the text and news headlines are visible to the viewers. The textual embeddings\footnote{Only posts having a minimum of 10 words are considered for this part of the analysis. This accounts for 5,405 (78.9\%) low-credibility and 15,883 (65.5\%) uncivil posts on Twitter, and 2,427 (58.6\%) low-credibility posts on FB.} are concatenated with individual's attributes, and time variable to get the final embedding of each post. \ahb{The confounders we choose to control for are based on prior literature~\cite{hasan2022impact,sahly2019social} and their feasibility of being measured. Apart from these confounders, there could also be certain other confounders (e.g., external events, algorithmic promotion effects, effect of ads) which we are unable to measure and thus account for in this study.} Our causal effect estimation has two steps: potential outcome prediction and matching. 

\noindent \textit{Potential outcome prediction}. The confounding in our case is time-varying and encoded in complex textual data, so we leverage the non-parametric nature of neural networks to learn deconfounded, low-dimensional representations of the high-dimensional data~\cite{koch_sainburg_geraldo_jiang_sun_foster_2021}. We leverage the Dragonnet\footnote{Dragonnet is chosen over other deep learning models for causal inference (S-learner, T-learner, TARNet) due to its ``Targeted Regularization'' procedure which allows for statistical guarantees~\cite{koch_sainburg_geraldo_jiang_sun_foster_2021}} model proposed by \citet{shi2019adapting} which uses feed-forward neural networks to learn balanced\footnote{Balancing is a treatment adjustment strategy that forces the treated and non-treated covariate distributions closer to deconfound the treatment from outcome \cite{johansson2016learning}} representations of the data such that each head models a separate potential outcome, a third propensity head predicts the propensity ($\pi$) of being treated and a free nudge parameter $\epsilon$ (see Appendix for model description). 

\begin{figure}[ht]
\setlength{\tabcolsep}{0pt}
\begin{tabular}{cc}

\includegraphics[width=0.5\linewidth]{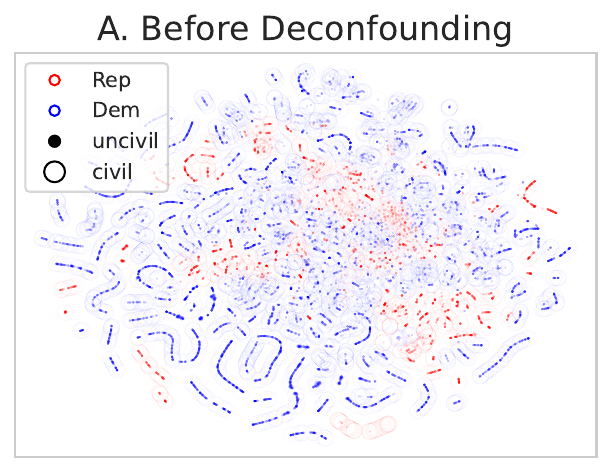}
&
\includegraphics[width=0.5\linewidth]{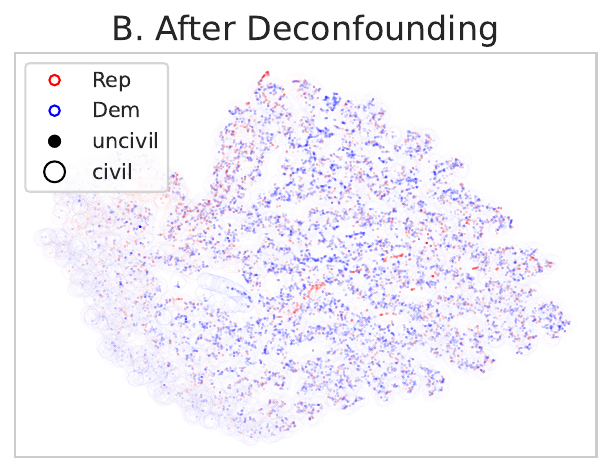}

\end{tabular}
\caption{{\bf Effectiveness of deconfounding for uncivil vs. civil tweets.} (A) shows the T-SNE space of content embeddings for civil vs. uncivil tweets by party. (B) shows the representation of the deconfounded embeddings returned by Dragonnet. After deconfounding, the representation space for treated and control covariate distributions ({\it party} in this example) can not be distinguished.} 
\label{fig:decon}
\end{figure}

We fine-tune the Dragonnet model\footnote{Models trained on a single NVIDIA A100 40GB PCIe GPU} by adding more hidden layers and using cross-entropy loss. The \textit{treatment variables} in our case are low-credibility and incivility respectively. We use a 5-fold cross-validation setting for training, with a 1:1 ratio of treated vs. non-treated random samples (see Appendix for model performance). \ahb{For low-credibility posts, we select corresponding non-treated posts that contain at least one URL and similarly include the URL headlines to minimize the confounding from text (e.g., presence of URL).} Figure~\ref{fig:decon} shows the effectiveness of our deconfounding for incivility on Twitter. 

\begin{figure}[ht]
\includegraphics[width=\linewidth]{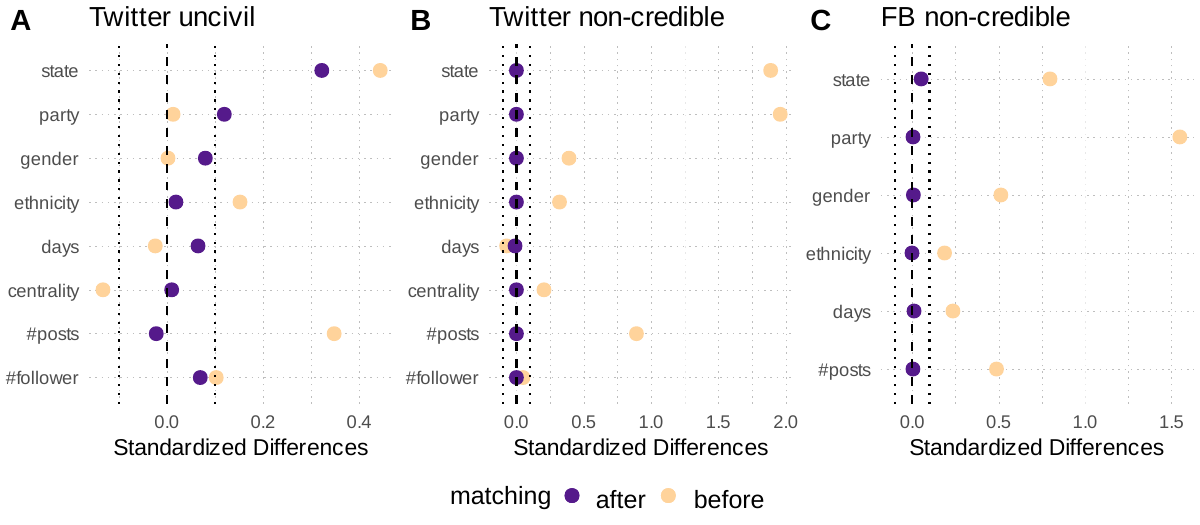}
\caption{{\bf Covariate balance after matching.} All covariates, except for party, and state in Twitter uncivil model, are balanced (i.e., absolute standardized difference $<0.1$ ) after matching on deconfounded embeddings.} \label{fig:match_bal}
\end{figure}

\noindent \textit{Matching.} For \textit{Conditional Average Treatment Effect} (CATE) estimation with Dragonnet predictions, we find that the covariates are not balanced after propensity score reweighting. To improve balance, we further use matching. We match the treated and untreated subjects based on the deconfounded Dragonnet embeddings (see Appendix). The covariate balance for matching is shown in Figure~\ref{fig:match_bal}. All the covariates are balanced for Twitter and FB low-credibility models. For Twitter incivility, all the covariates except for state and party are balanced after matching. The final CATE is calculated based on the matched samples as follows,

\begin{equation}
    CATE = \frac{1}{N^{\prime}} \sum\limits_{}^{N^{\prime}} (Y_i(1)-Y_{i^{\prime}}(0))
\end{equation}

where $Y(T)$ is the outcome for treatment $T$ and $N^{\prime}$ is the number of matched samples. We estimate the CATE for Democrats and Republicans separately to study potential asymmetries in receptivity across their audiences. Furthermore, we look at the CATE for ideologically extreme legislators to understand whether audiences engage differently with legislators at the extreme. We consider legislators having top 25\% conservative and top 25\% liberal ideological scores as Extreme Republicans and Extreme Democrats. 



\yrvn{Furthermore, our analysis ensures that outliers, a common occurrence in social networks, do not significantly impact our results. (See the Appendix for more details.)}

\section{Results}

\subsection{RQ1: Legislators' Visibility by Party, Gender, Ethnicity, Posting Frequency, State}

\begin{figure}[ht]
\includegraphics[width=\linewidth]{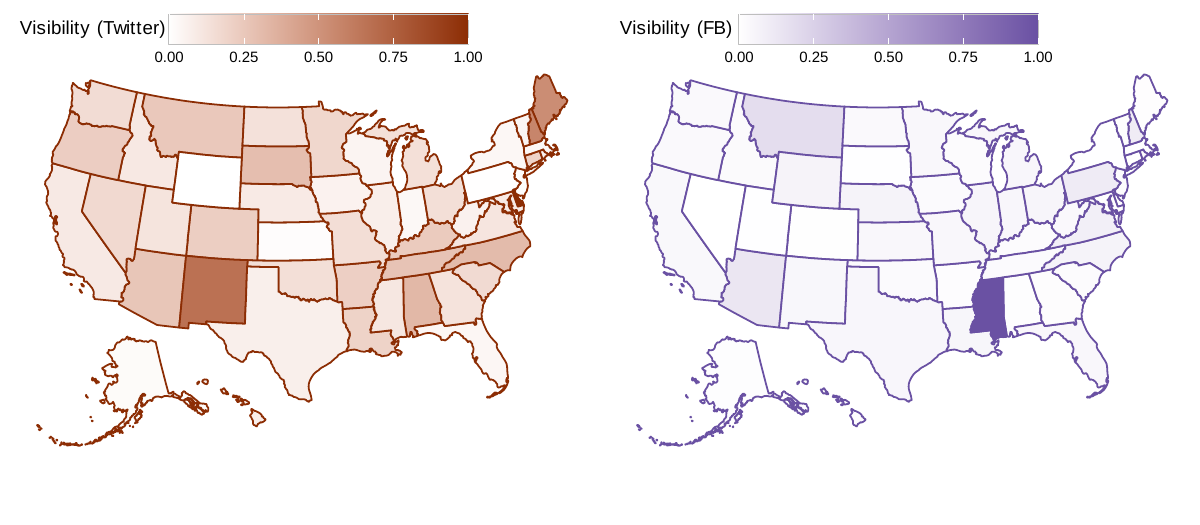}
\caption{{\bf Mean visibility of legislators from states.} The visibility ($V^{I_P}$) (normalized between 0-1) of legislators differ based on their state representation and platform.} \label{fig:map_vis}
\end{figure}

Table~\ref{tab:rq1_MWU} shows the effect sizes for the Mann-Whitney U test. We only report the results for $V^{I_P}$ here, the results for 25$th$, 50$th$, and 75$th$ percentile are added in the Appendix along with 95\% CI for Table~\ref{tab:rq1_MWU}. Overall, the visibility of legislators differs significantly based on party, gender, and posting frequency on both platforms and also for ethnicity on Twitter. Interestingly, Democrats and women appear to have higher visibility on Twitter but Republicans and men have higher visibility on FB ($p<0.001$). White legislators also receive more attention on Twitter. On both platforms, legislators with higher posting activity have higher visibility. Figure~\ref{fig:map_vis} shows the mean $V^{I_P}$ across US states for Twitter and FB. the visibility of legislators also differs based on their state representation. The most visible state on Twitter is New Mexico and Mississippi on FB. \ahb{The posting rate of legislators is the second highest for New Mexico on Twitter which could be a possible explanation for the high visibility\footnote{For instance on FB, Mississippi Republican senator Chris McDaniel has a remarkably high engagement which dominates the visibility term for the state.}.} 


We also look at the visibility of overlapping users (OL) on these platforms. Similar to prior results, Republicans and men receive higher engagement on FB and Democrats on Twitter. However, we do not find any significant difference across gender and ethnicity on Twitter for OL. Our results suggest that there exists cross-platform differences in how audiences engage with political content.


\begin{table}
\centering
\begin{threeparttable}
\setlength{\tabcolsep}{2pt}
\caption{RQ1 Effect sizes}\label{tab:rq1_MWU}
\small
\begin{tabular}{lcccc}
\toprule
& \multicolumn{2}{c}{} & \multicolumn{2}{c}{\textbf{Overlapping (OL)}} \\
 \cmidrule(l){4-5}
\textbf{IVs} & \textbf{Twitter} & \textbf{FB} & \textbf{Twitter} & \textbf{FB}\\
\cmidrule(r){1-1} \cmidrule(rl){2-3} \cmidrule(l){4-5}

Party & 0.239$^{***}$ & -0.299$^{***}$ & 0.189$^{***}$ & -0.347$^{***}$\\
\textit{(Rep vs. Dem)} & & & &\\
Gender & 0.076$^{***}$ & -0.089$^{***}$ & 0.009 & -0.128$^{***}$\\
\textit{(Men vs. Women)} & & & &\\
Ethnicity & -0.067$^{**}$ & -0.031 & -0.023 & -0.062\\
\textit{(White vs. Non-White)} & & & &\\
Posting Freq. & 0.457$^{***}$ & 0.435$^{***}$ & 0.368$^{***}$ & 0.418$^{***}$\\
\textit{($\leq$ vs. $>$ median)} & & & &\\

\bottomrule
\end{tabular}
\begin{tablenotes}
\small
\item $_{.}p < 0.1, ^{*}p < 0.05, ^{**}p < 0.01, ^{***}p < 0.001$
\end{tablenotes}
\end{threeparttable}
\end{table}

\subsection{RQ2: Factors Related to Visibility}

In RQ2 we look at factors related to legislators' platform visibility. The results for all DVs are similar but we only report results for $V^{I_P}$ in Table~\ref{tab:rq2_reg} (See Appendix for 95\% CI). We are unable to estimate the interaction effect between party and low-credibility posts for FB due to insufficient low-credibility posts from Democrats.
Party, gender, and post frequency are significantly correlated with visibility after controlling for other variables on both platforms. Republicans and men are more likely to garner higher visibility on FB whereas the opposite is true for Twitter. Higher posting activity is also related to higher visibility on both platforms. On Twitter, the number of followers and centrality in the intra-legislator network are not correlated with the legislators' visibility. Interestingly, there is a positive relation between legislators' network visibility and visibility.

As shown in Table~\ref{tab:rq2_reg}, the volume of low-credibility posts is positively associated with legislators' visibility on FB ($p<0.001$) but has an opposite effect ($p<0.01$) on Twitter. However, visibility is positively correlated with the interaction between party and low-credibility posts on Twitter, \ahb{i.e., one standard deviation increase in low-credibility posts by Republicans is associated with a 0.299 standard deviation increase in visibility}. Thus, posting low-credibility content is related to higher visibility for only Republicans, otherwise it has a negative association on Twitter. Posting more uncivil content also increases the visibility of legislators on Twitter ($p<0.001$). These results suggest that posting harmful content is associated with legislators' platform visibility.

We find similar results for OL, i.e., men and Republicans relate to higher visibility on FB, and Democrats on Twitter, again suggesting the cross-platform differences. Posting uncivil content on Twitter is positively associated with increased visibility, while posting low-credibility content is negatively associated with it. Moreover, visibility is positively associated with the interaction term between party and low-credibility posts. Interestingly however, posting low-credibility content is related to decreased visibility for the OL on FB similar to Twitter. Therefore, legislators who post content on both platforms have different communication strategies in comparison to the general legislator populations on those platforms which may be attributed to audience preferences across platforms. 

Thus, posting harmful content is related to the visibility of the legislators. But the observed correlation may be a {\it spurious} one due to confounders, such as content in similar topics. Next, we analyze whether incivility or low-credibility of a post {\it impacts} its visibility.





\begin{table}
\centering
\begin{threeparttable}
\setlength{\tabcolsep}{2pt}
\caption{RQ2 Regression Results}\label{tab:rq2_reg}
\small
\begin{tabular}{lcccc}
\toprule
& \multicolumn{2}{c}{} & \multicolumn{2}{c}{\textbf{Overlapping (OL)}} \\
 \cmidrule(l){4-5}
\textbf{IVs} & \textbf{Twitter} & \textbf{FB} & \textbf{Twitter} & \textbf{FB}\\
\cmidrule(r){1-1} \cmidrule(rl){2-3} \cmidrule(l){4-5}

Party [Rep]  & -0.133$^{**}$ & 0.423$^{***}$ & -0.142$^{*}$ & 0.512$^{***}$\\
Gender [Men]  & -0.078$^{*}$ & 0.090$^{**}$ & -0.011 & 0.113$^{**}$\\
Ethnicity [White]  & 0.020 & 0.015 & 0.020 & 0.040\\
\#posts  & 0.401$^{***}$ & 0.477$^{***}$ & 0.409$^{***}$ & 0.499$^{***}$\\
\#followers  & -0.028 & - & -0.028 & -\\
Centrality  & -0.020 & - & -0.029 & -\\
Network visibility  & 0.040$^{*}$ &  & 0.027 & -\\
\#\ahb{Low-Credible}  & -0.268$^{**}$ & 0.079$^{***}$ & -0.335$^{*}$ & -0.037$^{*}$\\
\#Uncivil  & 0.148$^{***}$ & - & 0.152$^{***}$ & -\\
Party [Rep] x & & & & \\
\#\ahb{Low-Credible} & 0.299$^{**}$ & - & 0.351$^{**}$ & -\\

\midrule
$R^2$ & 0.289 & 0.382 & 0.291 & 0.386\\

\bottomrule
\end{tabular}
\begin{tablenotes}
\small
\item $^{.}p < 0.1, ^{*}p < 0.05, ^{**}p < 0.01, ^{***}p < 0.001$
\end{tablenotes}
\end{threeparttable}
\end{table}

\subsection{RQ3: Observed Causal Impact of Incivility and Low-credibility on Visibility}

\begin{figure}[ht]
\includegraphics[width=\linewidth]{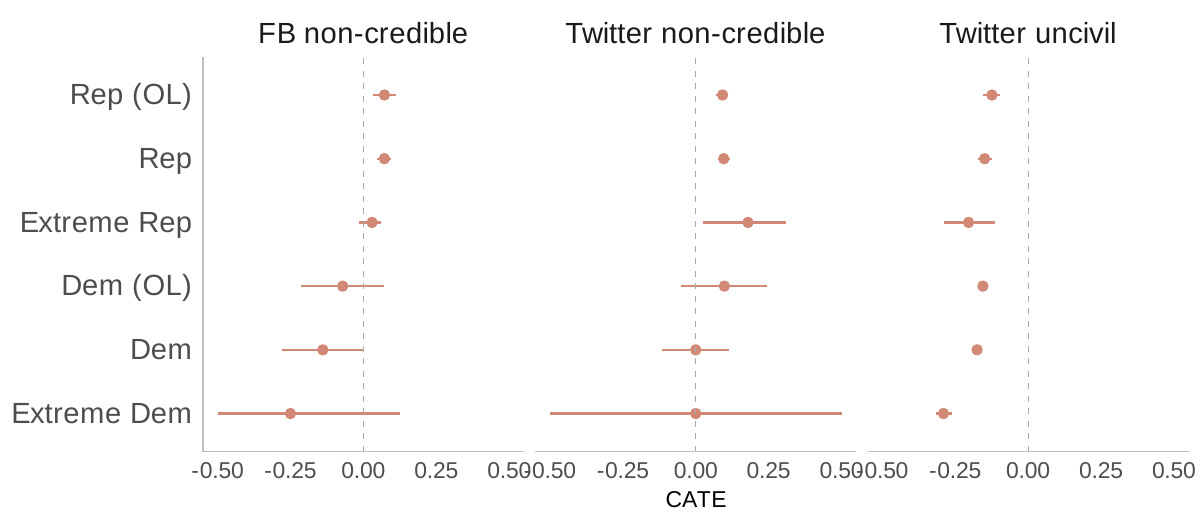}
\caption{{\bf Observed causal impact of low-credibility and incivility on legislators' content visibility.} After controlling for confounders, low-credibility has a positive impact on content visibility for Republicans on both platforms, but a negative effect for Democrats on FB. Incivility significantly hinders content visibility for all subgroups on Twitter.} \label{fig:ci_cate}
\end{figure}

Figure~\ref{fig:ci_cate} shows the CATE estimates and 95\% CI after matching. \ahbf{CATE is the expected change in the overperformance of a post if it contains low-credibility or incivility.} For FB, we find that Republican legislators receive higher attention on posting low-credibility content. Similar results hold when we look at Extreme and OL Republicans.
Interestingly, the visibility of Democrats decreases when they post low-credibility content on FB. There are no effects for Extreme and OL Democrat subgroups. 

For Twitter, similar to FB,  Republicans, including their OL and Extreme subgroups receive higher visibility on posting low-credibility content. The effect size is higher for Extreme Republicans, i.e., the more conservative a legislator is the higher attention they receive on posting misinformation. For Democrats, however, we do not find any significant effects. The difference in outcomes for Democrats across Twitter and FB could either be due to their distinct behaviors (e.g., content) and/or audience preferences across these platforms. The average number of posts containing low-credibility content by Democratic legislators is higher on Twitter than on FB as shown in Table~\ref{tab:mis_stats}. This may suggest that Democrats post less low-credibility content on FB since their visibility is penalized and/or vice-versa. 

The cross-platform analysis results suggest that there are asymmetries in how the Republican and Democratic party audiences engage with low-credibility content online and these asymmetries hold across platforms within and/or between parties. The former are more likely to engage with misinformation as shown by our results.

For uncivil content on Twitter, we find that both Democrats and Republicans, including their subgroups, receive lower engagement on posts containing uncivil language. The negative effects are higher for Democratic legislators compared to Republicans. The effects are also higher for Extreme legislators from both parties when compared to their party baselines. This implies that audiences irrespective of their partisan preferences engage less with uncivil content posted by legislators on Twitter and the legislators towards the extreme political spectra are penalized more.



\section{Discussion}
We analyzed the factors influencing the visibility dynamics of US state legislators by conducting a cross-platform analysis across Twitter and FB to understand different audience behaviors across these platforms. We showed that legislators' visibility varies based on their demographics, party, and posting frequency. Democrats have higher visibility on Twitter whereas Republicans have higher visibility on FB. Moreover, the regression analysis showed that a strong correlation exists between party and visibility, i.e., Democrats are associated with higher engagement on Twitter and Republicans on FB. These results also hold for the overlapping legislators, suggesting that audiences across these platforms engage with political content differently. Posting harmful content is also associated with legislators' visibility. Low-credibility content is related to increased visibility on FB, but decreased visibility on Twitter. Taking the effect of party into account, low-credibility posting correlates with higher visibility for Republicans. Uncivil posting is also correlated with higher account-level visibility on Twitter.

We further analyzed whether the increased visibility is due to posting harmful content after controlling for confounding factors such as demographics, party, topics, and time. Low-credibility garners more attention for Republicans on both platforms. However for Democrats,  low-credibility reduces their content visibility on Twitter. These results highlight the partisan asymmetries in how low-credibility content receives attention online. Existing works have shown population asymmetries~\cite{rao2022partisan}; our study further reveals attention disparity due to political actors' party affiliation. Higher online visibility provides a greater opportunity to influence public opinion (e.g., by gaining followers, impacting ranking algorithms). Politicians may post higher low-credibility content for political gains which may have implications for platform moderation.

Unlike low-credibility, incivility decreases the visibility of legislators' posts on Twitter for both Republicans and Democrats. The negative effects are more pronounced for Democrats compared to Republicans as well as for Extreme legislators, suggesting that audiences prefer to engage less with uncivil content irrespective of partisan preferences. Prior research has shown that uncivil content receives more negative reactions~\cite{coe2014online} owing to its emotionally charged language. The lower visibility may be attributed to the lack of expressing negative sentiments on Twitter, but further research is needed to confirm this. \ahb{Our results highlight the cross-platform differences as well as  asymmetries in how Democratic and Republican party audiences engage with political content online which is aligned with previous literature~\cite{kelm2020politicians,sahly2019social}}.

We show that harmful content is associated with legislators' online visibility. Such observed associations may be spurious, and other factors may contribute to their visibility. For instance, posting uncivil content has a positive association with visibility on Twitter but incivility has a negative impact on content visibility after controlling for confounders. Other factors may include the topics of their posts or simply the post timing. External factors (e.g., offline campaigns, media presence) can also contribute to their online visibility which is out of scope for this study. Moreover, there may be spillover effects from legislators' network visibility as hinted in our RQ2 results. Nevertheless, this study sheds light on some of the factors influencing legislators' overall as well as content visibility, but more research is needed to fully understand their online visibility dynamics.

\paragraph{Limitations and Future Work.} Our study has certain limitations. We only look at the years 2020 and 2021---the generalizability to other periods remains uncertain. Our method of identifying low-credibility content was conservative which could have led to certain biases in our sampling. We demonstrated the feasibility of addressing the representation biases in Section~\ref{methods}; however, it is uncertain whether we were able to fully correct for them. Furthermore, we only identified uncivil and low-credibility posts based on the textual content but do not consider other forms of content (e.g., images) which may also contain harmful information.

We only looked at the visibility based on total interactions received on posts without discriminating between positive and negative reactions. Future work can study the impact of harmful content on positive and negative visibility separately to get a more nuanced understanding of public receptivity. \ahb{It would also be interesting to look at the impact of posting harmful content on different types of engagement (e.g., Likes vs. Retweets).} Another possible extension could be adapting our causal study design for continuous treatment (e.g., how visibility is affected by the degree of incivility). 
\section{Ethical Considerations}
\paragraph{Data.} We collect data from two sources, Twitter and FB. For Twitter, the data was collected using Twitter's Official API v2.0 before rate limitations were imposed (i.e., March 2023). For FB, we collect data using CrowdTangle's official API by following their terms of service. All the data are publicly posted and available for viewing without restrictions.

We ensure that the Dragonnet model can effectively deconfound the covariate representation space for treated and non-treated samples based on our qualitative analysis and performance metrics. The classification errors from Dragonnet model are less likely to affect our results since we do not use the model predictions to calculate CATE. However, misclassification may still impact the deconfounding quality, resulting in confounding from textual content that we cannot measure, unlike other covariates.
Our study suggests that public figures sharing harmful content on social media has significant consequences. We showed that when low-credibility content is posted by public figures, the combination of user behavior (interacting with the posts) and platform mechanisms (e.g., feed recommendation algorithms) can result in increased visibility for such content. Our findings should not be interpreted as an encouragement to spread such low-credibility content; instead, they should serve as a warning that there may be incentives for political or elite actors to do so. Moreover, our study has focused on the behavior of US subnational politicians on two specific platforms---Twitter and FB. The results may not be generalizable to other platforms and social media users including the activities of political opinion leaders and media elites from other countries, or even US national politicians, due to several factors---media scrutiny, platform moderation rules, and public perception to name some. More research is needed to understand whether our results generalize to other settings.

\section*{Acknowledgement}
The authors would like to acknowledge support from AFOSR, ONR, Minerva, NSF \#2318461, and Pitt Cyber Institute's PCAG awards. The research was partly supported by Pitt's CRC resources (RRID:SCR 022735 through NIH \#S10OD028483). Any opinions, findings, and conclusions or recommendations expressed in this material do not necessarily reflect the views of the funding sources.

\bibliography{sections/main}



\clearpage
\section*{Appendix}

\begin{table*}
    \centering
    \small
    \setlength{\tabcolsep}{3pt}
    \renewcommand{\arraystretch}{1.3}
    \caption{Examples of Uncivil Posts on Twitter and FB, by party}\label{tab:uncivil_ex}
    \begin{tabular}{|c|c|p{14cm}|}
        \hline
        {\bf platform} & {\bf party} & {\bf text}\\
        \hline
        FB & Rep & {\it While millions of Americans have yet to receive their stimulus checks, a new study reveals that \$4.38 billion of the new round will go right into the pockets of illegal immigrants. Another dumb idea and stupid stupid policy. What is wrong with these people?}\\
        FB & Dem & {\it ``I didn’t think it would be this ridiculous. It’s embarrassing to be a state senator at this point,'' Paul Boyer said of partisan recount. Yes it does make you look like idiots. Wasting time \& resources on \#TrumpsBigLie}\\
        Twitter & Rep & {\it We are at the start of one of the LARGEST recessions in American history, which will DESTROY many lives, and people are still in favor of partial lockdownshow stupid could you possibly be! Bunch of damn fools.}\\
        Twitter & Dem & {\it You are a blithering idiot. Who gives a shit about the VP. Vote for Trump and thousands upon thousands will die.}\\
        \hline
    \end{tabular}
\end{table*}

\paragraph{Data Collection:} \ahb{For Twitter data, we employed a comprehensive approach, drawing from various sources, including existing datasets~\cite{kim2022attention}, and conducting searches on Google, Twitter, Wikipedia, legislators' official websites, campaign sites, and Ballotpedia, to compile the accounts and demographic information of state legislators. This meticulous strategy facilitated the manual identification and verification of legislators' Twitter accounts. Subsequently, we refined our dataset to only include legislators who served between 2020 and 2021, determined by their tenure in office. It is important to acknowledge that the completeness of our data was affected by factors such as inactive or inaccessible accounts after legislators left office.}

\ahb{Our initial approach to collecting FB data involved gathering posts from accounts bearing names indicative of belonging to state legislators. Subsequent verification against information from Ballotpedia allowed us to filter out non-legislator accounts. To address data gaps, we conducted three successive rounds of data recollection in April 2022, March 2023, and April 2023. The successive rounds allowed us to capture posts from accounts previously overlooked. However, similar to Twitter data, numerous accounts had become inaccessible, largely due to campaign or official accounts no longer being active.}

\ahb{Despite these efforts, we encountered a significant challenge with FB data collection concerning accounts that were not listed on Ballotpedia. Although we attempted to identify missing accounts using keyword searches, achieving a perfect match with the legislator information on Ballotpedia was difficult. This limitation resulted in a higher rate of mismatch in the mapped attributes of FB accounts, i.e., around 70.2\% of legislators could be mapped to their attributes for FB. Table~\ref{tab:fb_stats} shows the statistics of our FB dataset after mapping the legislators to their attributes. The trends are similar to that in Table 2 which suggests that no or minimal biases were introduced during our mapping process.}

\begin{table}
\centering
\setlength{\tabcolsep}{3pt}
\caption{Descriptive statistics for FB dataset after mapping, showing the number of legislators, posts, and median interactions received per post by party.}\label{tab:fb_stats}
\small
\begin{tabular}{lccc}
\toprule
 & \multicolumn{3}{c}{\textbf{FB}} \\
\cmidrule(rl){2-4} 
\textbf{party} & \textbf{\#users} & \textbf{\#tweets} & \textbf{Int} \\
\cmidrule(r){1-1} \cmidrule(rl){2-4}
Dem & 1588 & 171K & 61.0\\
Rep & 1718 & 152K & 114.0\\
\bottomrule
\end{tabular}
\end{table}

\begin{figure}[ht]
\centering
\includegraphics[width=1\linewidth]{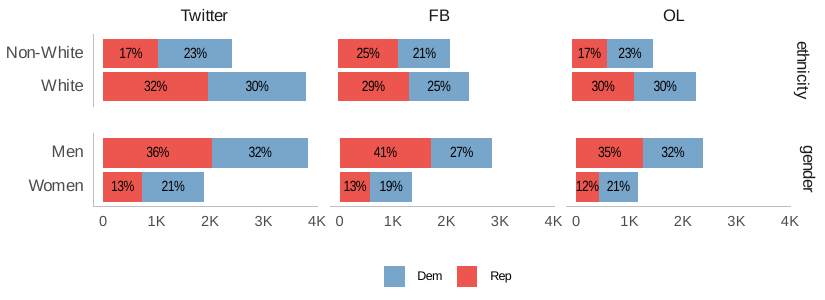}
\caption{Ethnicity and gender by party, platform, and for overlapping users. Our dataset has higher representation of men and White legislators on both platforms.} \label{fig:att}
\end{figure}

Figure~\ref{fig:att} shows the breakdown of ethnicity and gender by party and for OL. 

\paragraph{Assessing Post's Civility.} \ahb{
For a study like ours it is hard to interpret the continuous toxicity scores returned by the Detoxify model. So we follow the common practice in literature to convert the toxicity scores to binary based on a threshold~\cite{hua2020characterizing}. It is important to have a threshold for the toxicity for our particular dataset because uncivil language may evolve over different topics and time.} To estimate an ideal cutoff for the toxicity score, we manually annotate a sample of 300 posts as uncivil or civil using stratified sampling such that more samples are included at higher toxicity scores. Three annotators labeled all 300 samples based on the aforementioned definition of toxic language. Since Cohen's Kappa suffers from imbalanced label distributions, we compute the pairwise agreement scores for the percentage of total samples agreed upon by the annotators, which ranges between 66.3-85.3\%. The final labels are decided according to the majority vote. Based on the ROC (AUC $=$ 0.81), we choose the cutoff for toxicity as 0.82, i.e., posts having a score above 0.82  are considered uncivil. \ahb{This is similar to previous works using Detoxify or Perspective API which have a threshold between 0.5-0.9~\cite{hua2020characterizing}, though we acknowledge that our threshold is more towards the conservative side which is done to ensure that uncivil posts are selected with high precision.} Table~\ref{tab:uncivil_ex} shows examples of uncivil posts by legislators on Twitter and FB.

\ahb{The number of uncivil posts on FB\footnote{The number of uncivil posts is still low at other cutoffs, for e.g., cutoff = 0.1 yields around 2,690 uncivil posts} is much lesser compared to Twitter which shows that the political communication styles are different on two platforms. This finding also resonates with previous studies which have suggested that FB is used more for broadcasting purposes whereas Twitter is used more for direct communication~\cite{enli2013personalized}. Based on this, it is reasonable that there are very less uncivil posts on FB since the language used on FB tends to be more formal. 
We further verify this by measuring the readability scores of author’s posts on these two platforms using Flesch–Kincaid grade level. The median readability of legislators is 11.04 (i.e., the text is written at a level suitable for someone who has completed the 11th grade in the US education system) on FB and 9.55 on Twitter. This shows that the language used on FB is indeed more formal and potentially the reason why it is more civil.}

\paragraph{Measuring Visibility.} \ahb{The visibility metric is designed to capture the overall engagement on the platforms and hence our metric includes all/most of the elements used to calculate engagement on the respective platforms. We further analyze the contribution of individual visibility elements on each platform. On Twitter, Likes contribute 2.3\% and Retweets contribute 97.7\% to the overall interactions. Reply and quote consist of a small percentage of the interactions. On FB however, Likes contribute 51.8\%, Shares contribute 17.0\% and Comments contribute 13.2\% to the overall interactions. This suggests that audiences engage differently with content on Twitter and FB, e.g., retweeting is most dominant form of engagement on Twitter whereas Liking for FB. Moreover, the interpretation of individual elements may also be different across platforms, for instance, audiences may engage with Like on Twitter differently than Like on FB simply due to different icons, therefore a direct comparison of individual engagement metrics may not be justified. So, by only including individual elements, the visibility metric may not be able to capture the level of engagement on these platforms, making the interpretation of the metric harder for a cross-platform study.}

\ahb{We further examine the possibility of the visibility metric being dominated by a single element by testing if the underlying distributions are similar for the total interactions and the most dominant element. According to our KS tests, for both Likes on FB and Retweets on Twitter, we find that the distributions are significantly different (p-value$<$0.05) compared to the total interactions, suggesting that other elements also contribute to the overall engagement on both platforms and hence need to be included in the visibility measure.}

\paragraph{DV Transformation for RQ2.} To satisfy assumptions of linear regression in RQ2, we transform all variables to be close to normal distributions using the ``bestNormalize'' R package. \ahb{For Twitter, we transform variables as follows: $V^{I_P}$ (Yeo-Johnson), \#posts (Box Cox), \#Misinfo (sqrt), \#Uncivil (sqrt), Network visibility (Center+scale), \#followers (Box Cox), Centrality (None). For FB, we transform variables as follows: $V^{I_P}$ (Yeo-Johnson), \#posts (Box Cox), \#Misinfo (sqrt).}    

\begin{table}
\vspace{-1em}
\centering
\setlength{\tabcolsep}{1pt}
\caption{95\% CI for Table 4}\label{tab:rq1_mwu_ci}
\scriptsize
\begin{tabular}{lcccc}
\toprule
& \multicolumn{2}{c}{} & \multicolumn{2}{c}{\textbf{OL}} \\
 \cmidrule(l){4-5}
\textbf{IVs} & \textbf{Twitter [CI]} & \textbf{FB [CI]} & \textbf{Twitter [CI]} & \textbf{FB [CI]}\\
\cmidrule(r){1-1} \cmidrule(rl){2-3} \cmidrule(l){4-5}


Party & 0.239 & -0.299 & 0.189 & -0.347\\
 & [0.202, 0.277] & [-0.337, -0.259] & [0.141, 0.240] & [-0.397, -0.298]\\
Gender & 0.076 & -0.089 & 0.009 & -0.128\\
 & [0.033, 0.117] & [-0.132, -0.046] & [-0.044, 0.059] & [-0.182, -0.072]\\
Ethnicity & -0.067 & -0.031 & -0.023 & -0.062\\
 & [-0.108, -0.027] & [-0.079, 0.015] & [-0.073,  0.029] & [-0.128, 0.001]\\
Posting & 0.457 & 0.435 & 0.368 & 0.418\\
Freq. & [0.427, 0.492] & [0.398, 0.470] & [0.323, 0.414] & [0.370, 0.461]\\

\bottomrule
\end{tabular}
\vspace{-1.4em}
\end{table}

\paragraph{RQ1 Tables.} Table~\ref{tab:rq1_mwu_ci} shows the 95\% CI for Table 4. Tables~\ref{tab:rq1_tw_rob} and ~\ref{tab:rq1_fb_rob} show the results for 25$th$, 50$th$, and 75$th$ percentile visibility for Twitter and FB respectively.





\begin{table}
\vspace{-1.5em}
\centering
\setlength{\tabcolsep}{1pt}
\caption{Robustness analysis of RQ1 results for 25$th$, 50$th$, and 75$th$ percentile visibility on Twitter}\label{tab:rq1_tw_rob}
\scriptsize
\begin{threeparttable}
\begin{tabular}{lccc}
\toprule

\textbf{IVs} & \textbf{$25^{th}$} & \textbf{$50^{th}$} & \textbf{$75^{th}$}\\
\cmidrule(r){1-1} \cmidrule(rl){2-4}

Party & 0.120$^{***}$ & 0.160$^{***}$ & 0.160$^{***}$\\
Gender & 0.016 & 0.036$_{.}$ & 0.060$^{**}$\\
Ethnicity & 0.004 & -0.027 & -0.054$^{**}$\\
Posting Freq. & 0.255$^{***}$ & 0.354$^{***}$ & 0.386$^{***}$\\

\bottomrule
\end{tabular}
\end{threeparttable}
\end{table}

\begin{table}
\vspace{-1em}
\centering
\setlength{\tabcolsep}{1pt}
\caption{Robustness analysis of RQ1 results for 25$th$, 50$th$, and 75$th$ percentile visibility on FB}\label{tab:rq1_fb_rob}
\scriptsize
\begin{threeparttable}
\begin{tabular}{lccc}
\toprule

\textbf{IVs} & \textbf{$25^{th}$} & \textbf{$50^{th}$} & \textbf{$75^{th}$}\\
\cmidrule(r){1-1} \cmidrule(rl){2-4}

Party & -0.293$^{***}$ & -0.300$^{***}$ & -0.291$^{***}$\\
Gender & -0.078$^{***}$ & -0.088$^{***}$ & -0.088$^{***}$\\
Ethnicity & -0.048 & -0.047$_{.}$ & -0.032\\
Posting Freq. & 0.376$^{***}$ & 0.402$^{***}$ & 0.424$^{***}$\\

\bottomrule
\end{tabular}
\end{threeparttable}
\vspace{-1.5em}
\end{table}

\paragraph{RQ2 Tables.} Table~\ref{tab:rq2_reg_ci} shows the 95\% CI for Table 5.

\begin{figure}[ht]
\centering
\includegraphics[width=0.75\linewidth]{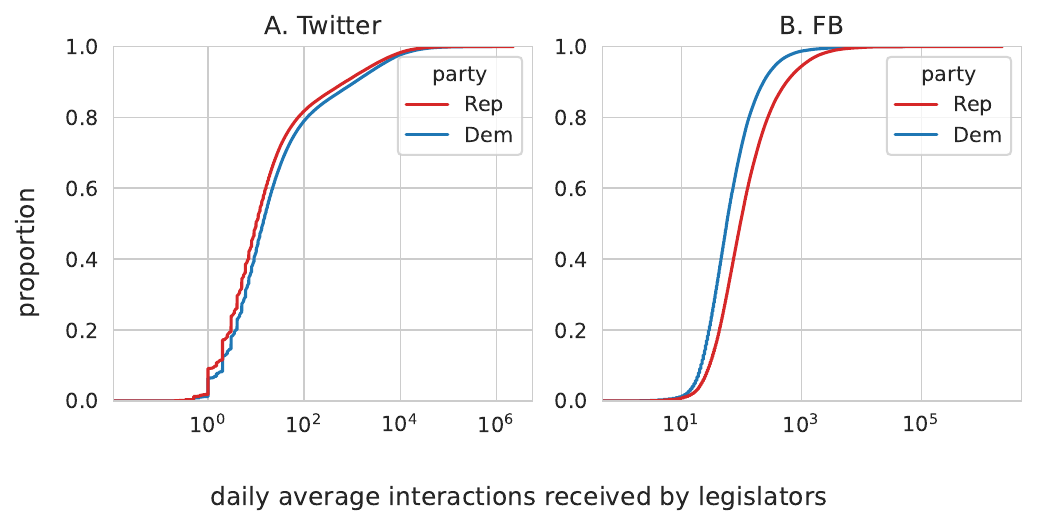}
\caption{ECDF plots for daily mean interactions received by legislators on Twitter and FB, by party.} \label{fig:b}
\end{figure}

\paragraph{Benchmarks for \ahb{$thres$}, $w$.}\label{bench}
Figure~\ref{fig:b} shows the ECDF plots for daily mean interactions received by legislators on Twitter and FB, by party. For Twitter and FB, the medians are close to 10 and 100 respectively. So we select \ahb{$thres=10$} for Twitter and \ahb{$thres=100$} for FB. We do not have different \ahb{$thres$} across parties since medians are similar across parties on both platforms.  

\begin{figure}[ht]
\centering
\includegraphics[width=0.75\linewidth]{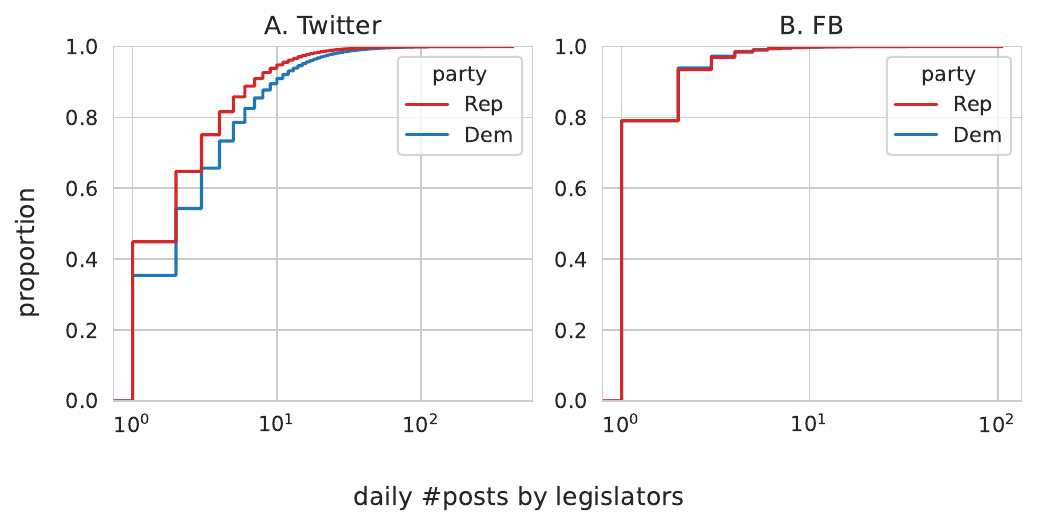}
\caption{ECDF plots for daily number of posts by legislators on Twitter and FB, by party.} \label{fig:w}
\vspace{-1.2em}
\end{figure}

Figure~\ref{fig:w} shows the ECDF plots for daily number of posts by legislators on Twitter and FB, by party. Again the median posting rates are similar for Republicans and Democrats on both Twitter and FB. The median daily post count on Twitter is 2 and 1 on FB. To ensure that we have a sufficient number of posts per legislator to compute the overperforming scores and simultaneously account for temporal variation in our data, we select $w=14$ for both Twitter and FB.

\paragraph{Dragonnet Model Description.} Dragonnet uses feed-forward neural networks to learn balanced representations of the data such that each head models a separate potential outcome, a third propensity head predicts the propensity ($\pi$) of being treated, and a free nudge parameter $\epsilon$. The $\pi$ and $\epsilon$ parameters are used to re-weight the outcomes to provide lower biased estimates of the \textit{Conditional Average Treatment Effect} (CATE). The error gradients from the two outcome modeling heads are propagated back to the shared representation layers of the Dragonnet model to learn the covariate representation, i.e., $\phi(X)$. The representation layers learn a balanced representation of the data since the model objective is to predict both outcomes and each outcome modeling head learns a function of the transformed covariate representation, i.e., $Y(T)=h(\phi(x),T)$. The CATE from Dragonnet predictions is estimated as follows,

\begin{equation}
    CATE = \frac{1}{N} \sum\limits_{i}^{N} (Y^{*}_i(1)-Y^{*}_i(0))
\end{equation}

where,
\begin{equation}
Y^{*}_i = \hat{Y}_i + (\frac{T_i}{\pi(\phi(X_i), 1)}-\frac{1-T_i}{\pi(\phi(X_i), 0)}) \times \epsilon
\end{equation}

where $\hat{Y}(1)$ and $\hat{Y}(0)$ are predictions returned by the two outcome modeling heads respectively, $\pi$ is the predicted propensity of a sample being treated, and sample size $N$. 

\begin{table}
\centering
\setlength{\tabcolsep}{1pt}
\caption{Dragonnet performance}\label{tab:dr_auc}
\scriptsize
\begin{tabular}{lcccc}
\toprule
& \multicolumn{2}{c}{\textbf{AUC}} & \multicolumn{2}{c}{\textbf{Macro F1}} \\
 \cmidrule(l){2-3} \cmidrule(l){4-5}
 & \textbf{overall} & \textbf{Extreme} & \textbf{overall} & \textbf{Extreme}\\
\cmidrule(r){1-1} \cmidrule(rl){2-3} \cmidrule(l){4-5}

Twitter uncivil & 0.74 & 0.72 & 0.69 & 0.66 \\
Twitter non-credible & 0.73 & 0.74 & 0.66 & 0.68 \\
FB non-credible & 0.80 & 0.88 & 0.74 & 0.82 \\

\bottomrule
\end{tabular}
\end{table}

\begin{figure}[ht]
\centering
\setlength{\tabcolsep}{0pt}
\begin{tabular}{ccc}
\includegraphics[width=.33\linewidth]{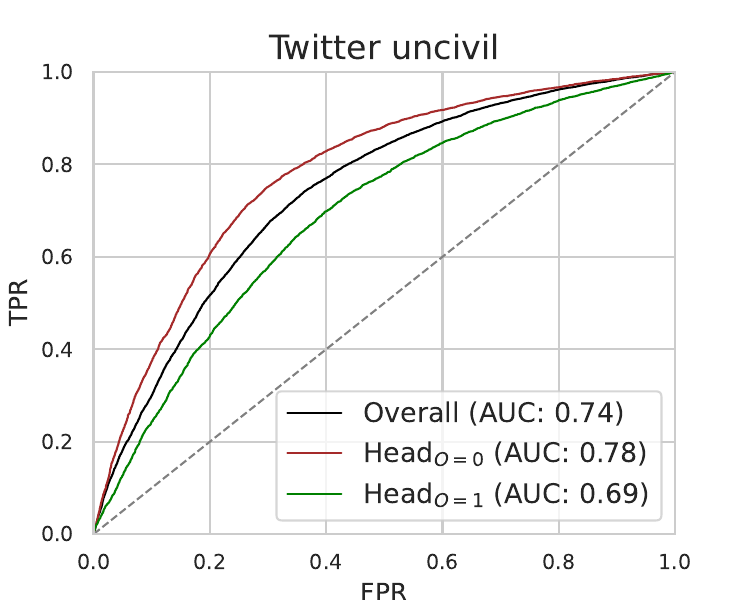}
&
\includegraphics[width=.33\linewidth]{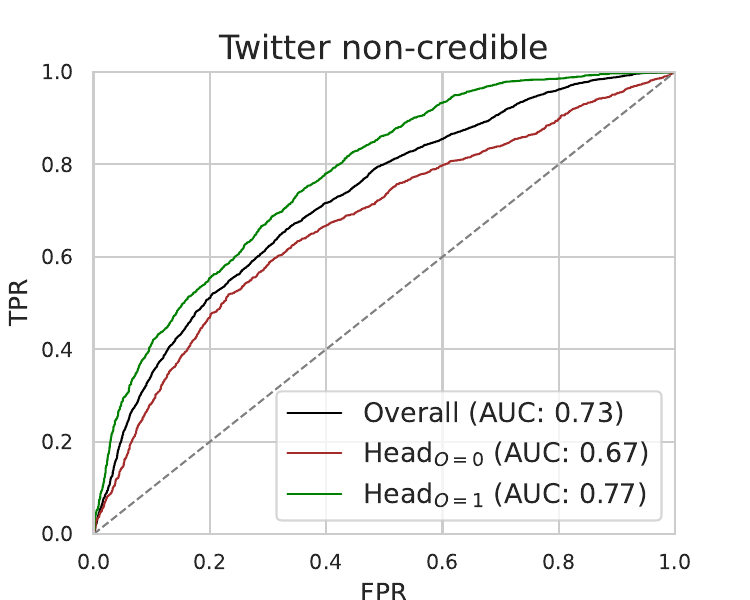} 
&
{\includegraphics[width=.33\linewidth]{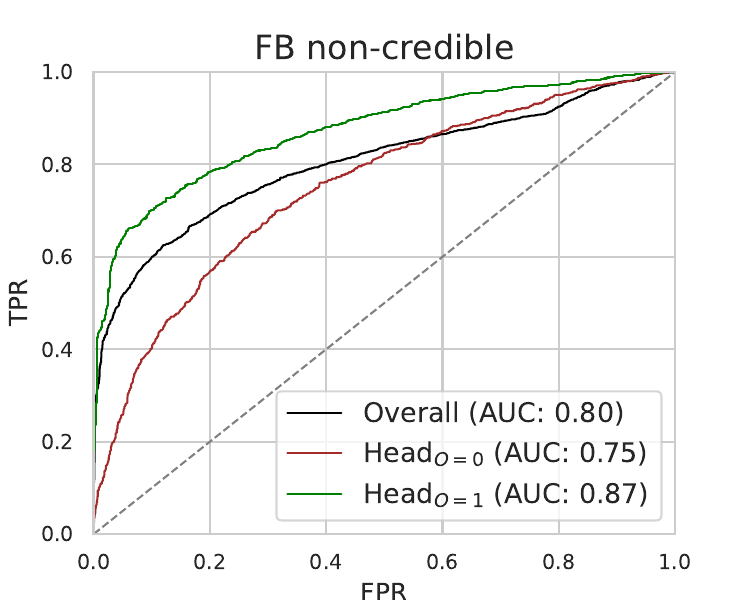}}
\end{tabular}
\vspace{1em}
\caption{ROC curves showing Dragonnet performance of overall as well as two outcome modeling heads, for Twitter incivility, Twitter and FB low-credibility respectively.} 
\label{fig:roc}
\end{figure}

\paragraph{Dragonnet Model Performance.}\label{dr}
Figure~\ref{fig:roc} shows the ROC curves for Twitter incivility, Twitter low-credibility and FB low-credibility Dragonnet models. The AUC and Macro F1-scores\footnote{F1-scores reported at optimal cutoff} are reported in Table~\ref{tab:dr_auc}.

\begin{table}
\centering
\setlength{\tabcolsep}{0pt}
\caption{95\% CI for Table 5}\label{tab:rq2_reg_ci}
\scriptsize
\begin{tabular}{lcccc}
\toprule
& \multicolumn{2}{c}{} & \multicolumn{2}{c}{\textbf{OL}} \\
 \cmidrule(l){4-5}
\textbf{IVs} & \textbf{Twitter} & \textbf{FB} & \textbf{Twitter} & \textbf{FB}\\
\cmidrule(r){1-1} \cmidrule(rl){2-3} \cmidrule(l){4-5}

Party [Rep]  & -0.133 & 0.423 & -0.142 & 0.512\\
 & [-0.219, -0.046] & [0.359, 0.487] & [-0.252, -0.029] & [0.431, 0.594] \\
Gender [Men]  & -0.078 & 0.090 & -0.011 & 0.113\\
 & [-0.149, -0.007] & [0.029, 0.152] & [-0.101, 0.079] & [0.033, 0.194]\\
Ethnicity [White]  & 0.020 & 0.015 & 0.020 & 0.040\\
 & [-0.061, 0.100] & [-0.056, 0.086] & [-0.079, 0.118] & [-0.054, 0.133]\\
\#posts  & 0.401 & 0.477 & 0.409 & 0.499\\
 & [0.345, 0.455] & [0.443, 0.510] & [0.337, 0.481] & [0.459, 0.538]\\
\#followers  & -0.028 & - & -0.028 & -\\
 & [-0.076, 0.021] & & [-0.091, 0.034] & \\
Centrality  & -0.020 & - & -0.029 & -\\
 & [-0.060, 0.019] & & [-0.079, 0.021] & \\
Network visibility  & 0.040 &  & 0.027 & -\\
 & [0.002, 0.080] & & [-0.020, 0.076] & \\
\#\ahb{Low-Credible}  & -0.268 & 0.079 & -0.335 & -0.037\\
 & [-0.466, -0.070] & [0.047, 0.110] & [-0.601, -0.069] & [-0.074, 0.000]\\
\#Uncivil  & 0.148 & - & 0.152 & -\\
 & [0.101, 0.196] & & [0.092, 0.213] & \\
 Party [Rep] x  & 0.299& - & 0.351 & -\\
\#\ahb{Low-Credible} & [0.103, 0.496] & & [0.087, 0.614] & \\

\midrule
$R^2$ & 0.289 & 0.382 & 0.291 & 0.386\\

\bottomrule
\end{tabular}
\vspace{-1.5em}
\end{table}

\paragraph{Covariate Balance for Matching.}\label{bal}
We employ 1:1 matching such that each treated sample is matched to one untreated sample. We use Nearest Neighbour matching based on Euclidean distance between the deconfounded tweet embeddings. We find matches for 9677 (64.0\%) uncivil tweets, 3957 (73.5\%) low-credibility tweets, and 1583 (61.1\%) low-credibility FB posts using a distance cutoff of 0.1 to maximize the covariate overlap. This gives us balanced representations of the observed covariates across the treated and untreated samples as shown in Figure 6. We compute the standardized differences for each of the covariates before and after matching. A score between -0.1-0.1 indicates balance.

\begin{figure}[ht]
\centering
\setlength{\tabcolsep}{0pt}
\begin{tabular}{ccc}
\includegraphics[width=.5\linewidth]{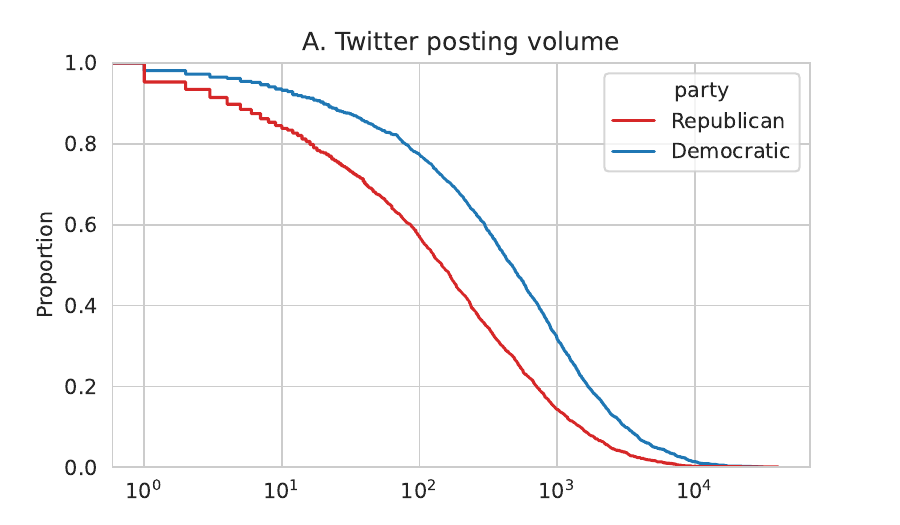}

&
{\includegraphics[width=.5\linewidth]{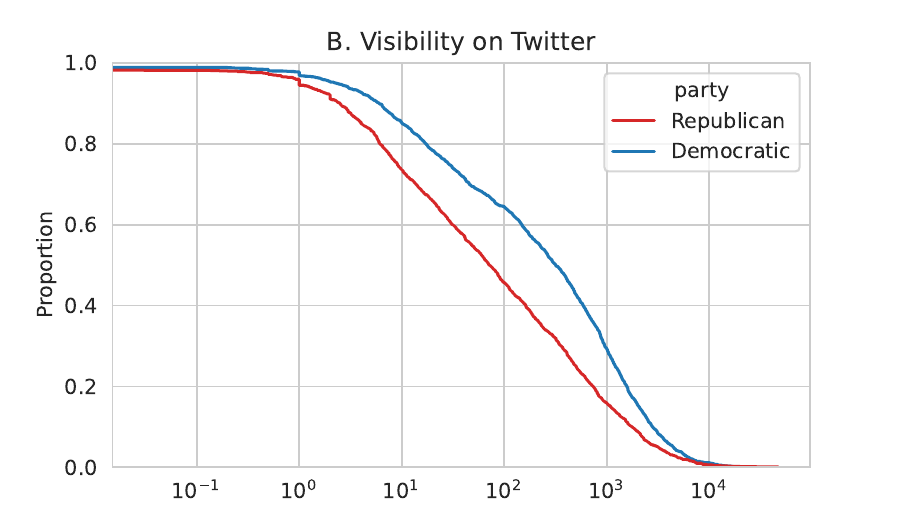}}
\end{tabular}
\vspace{1em}
\caption{Scale-free distributions for (A) Posting volumes and (B) Visibility of legislators on Twitter. The distributions are also similar for FB.} 
\label{fig:powerLaw}
\end{figure}

\paragraph{Effect of Outliers}
\yrvn{Outliers are common in social networks, but their impact on analysis results and conclusions can vary. In our dataset, outliers may exist in terms of posting volume and engagement received due to the scale-free distributions (refer to Fig~\ref{fig:powerLaw}). However, we have taken measures to ensure these outliers do not significantly impact our results.}

\yrvn{For RQ1, we used a non-parametric statistical test, the Mann-Whitney U test, to compare distributions, which is robust to outliers. For RQ2, we used non-linear transformations on study variables to minimize the impact of outliers in the regression analysis.}

\yrvn{In RQ3, we matched accounts with similar characteristics in the de-confounded embedding space, such as similar posting rates. This process either discarded outliers if no adequate match was found or retained them if an adequate match was identified. This matching step ensured the balance of the covariates before running the estimation of the Conditional Average Treatment Effect (CATE), further reducing the impact of outliers.}

\end{document}